\documentclass[longauth,letter]{aa}  
\usepackage{graphicx}
\usepackage{natbib}

\newcommand{\sub}[1]{_{\rm #1}}

\newcommand{\CII}{[C\,{\sc ii}]}
\newcommand{\CI}{[C\,{\sc i}]}
\newcommand{\HII}{H\,{\sc ii}}
\newcommand{\HI}{H\,{\sc i}}
\newcommand{\OI}{[O\,{\sc i}]}
\newcommand{\kms}{km~s$^{-1}$}
\newcommand{\changed}{}

\usepackage{txfonts}
\begin{document}
\title{HIFI observations of warm gas in DR21: Shock versus radiative 
heating\thanks{{\it Herschel} is an ESA space observatory with 
science instruments 
provided by European-led Principal Investigator consortia and with 
important participation from NASA.}}
\author{V.~Ossenkopf\inst{1,2}$\!$, M.~R\"ollig\inst{1}, R.~Simon\inst{1}, 
N.~Schneider\inst{3}, Y.~Okada\inst{1}, J.~Stutzki\inst{1}, M.~Gerin\inst{4},
M.~Akyilmaz\inst{1}, D.~Beintema\inst{2}, A.O.~Benz\inst{5}, O.~Berne\inst{6},
F.~Boulanger\inst{7}, B.~Bumble\inst{8},
O.~Coeur-Joly\inst{9,10}, C.~Dedes\inst{5}, M.C.~Diez-Gonzalez\inst{11},
K.~France\inst{12},
A.~Fuente\inst{13},  J.D.~Gallego\inst{11}, J.R.~Goicoechea \inst{14},
R.~G\"usten\inst{15}, A.~Harris\inst{16}, R.~Higgins\inst{171}, 
B.~Jackson\inst{2}, C.~Jarchow\inst{18},
C.~Joblin\inst{9,10}, T.~Klein\inst{15}, C.~Kramer\inst{19},
S.~Lord\inst{20}, P.~Martin\inst{12}, J.~Martin-Pintado\inst{14},
B.~Mookerjea\inst{21}, D.A.~Neufeld\inst{22}, T.~Phillips\inst{23}, J.R.~Rizzo\inst{14},
F.F.S.~van der Tak\inst{2,24},
D.~Teyssier\inst{25}, H.~Yorke\inst{8}
}

\institute{I. Physikalisches Institut der Universit\"at 
zu K\"oln, Z\"ulpicher Stra\ss{}e 77, 50937 K\"oln, Germany
\and
SRON Netherlands Institute for Space Research, P.O. Box 800, 9700 AV 
Groningen, Netherlands
\and
Laboratoire AIM, CEA/DSM - INSU/CNRS - Universit\'e Paris Diderot, IRFU/SAp CEA-Saclay, 91191 Gif-surYvette, France
\and
LERMA \& UMR 8112 du CNRS, Observatoire de Paris and \'Ecole Normale Sup\'erieure, 24 Rue Lhomond, 75231 Paris Cedex 05, France
\and
Institute for Astronomy, ETH Z\"urich, 8093 Z\"urich, Switzerland
\and
Leiden Observatory, Universiteit Leiden, P.O. Box 9513, NL-2300 RA Leiden, The Netherlands 
\and
Institut d'Astrophysique Spatiale, Universit\'e Paris-Sud, B\^at. 121, 91405 Orsay Cedex, France
\and
Jet Propulsion Laboratory, 4800 Oak Grove Drive, MC 302-231, Pasadena, CA 91109  U.S.A.
\and
Universit\'e de Toulouse, UPS, CESR, 9 avenue du colonel Roche, 31028 Toulouse cedex 4, France
\and
CNRS, UMR 5187, 31028 Toulouse, France
\and
Observatorio Astron\'omico Nacional (IGN), Centro Astron\'omico de Yebes, Apartado 148. 19080 Guadalajara
\and
Department of Astronomy and Astrophysics, University of Toronto, 60 St. George Street, Toronto, ON M5S 3H8, Canada
\and
Observatorio Astron\'omico Nacional (OAN), Apdo. 112, 28803 Alcal\'a de Henares (Madrid), Spain
\and
Centro de Astrobiolog\'ia, CSIC-INTA, 28850, Madrid, Spain
\and
Max-Planck-Institut f\"ur Radioastronomie, Auf dem H\"ugel 69, 53121, Bonn, Germany
\and
Astronomy Department, University of Maryland, College Park, MD 20742, USA
\and
Experimental Physics Dept., National University of Ireland: Maynooth, Co. Kildare, Ireland
\and
MPI f\"ur Sonnensystemforschung, D 37191 Katlenburg-Lindau, Germany
\and
Instituto de Radio Astronom\'ia Milim\'etrica (IRAM), Avenida Divina Pastora 7, Local 20, 18012 Granada, Spain
\and
IPAC/Caltech, MS 100-22, Pasadena, CA 91125, USA
\and 
Tata Institute of Fundamental Research (TIFR), Homi Bhabha Road, Mumbai 400005, India
\and 
Department of Physics and Astronomy, Johns Hopkins University, 3400 North Charles Street, Baltimore, MD 21218, USA
\and
California Institute of Technology, 320-47, Pasadena, CA  91125-4700, USA
\and
Kapteyn Astronomical Institute, University of Groningen, P.O. Box 800, 9700 AV Groningen, Netherlands
\and
European Space Astronomy Centre, Urb. Villafranca del Castillo, P.O. Box 50727, Madrid 28080, Spain
}

\authorrunning {V.~Ossenkopf, M.~R\"ollig, R.~Simon\inst{1} et al.} 
\titlerunning{HIFI observations of warm gas in DR21}

\abstract
{
The molecular gas in the DR21 massive star formation region is 
known to be affected by the strong UV field from the central star cluster
and by a fast outflow creating a bright shock. The relative contribution
of both heating mechanisms is the matter of a long debate.
}
{By better sampling the excitation ladder of various tracers
we provide a quantitative distinction between the different heating mechanisms.
}
{
HIFI observations of mid-$J$ transitions of CO and HCO$^+$ isotopes
allow us to bridge the gap in excitation energies between observations
from the ground, 
characterizing the cooler gas, and existing ISO LWS spectra, constraining
the properties of the hot gas. Comparing the detailed line profiles
allows to identify the physical structure of the different components. 
}
{
In spite of the known shock-excitation of H$_2$ and the clearly
visible strong outflow, we find that the emission of all lines
up to $\ga2$~THz can be explained by purely
radiative heating of the material. However, the new \textit{Herschel/HIFI}
observations reveal two types of excitation conditions. We find hot 
and dense clumps close to the central cluster, probably dynamically
affected by the outflow, and a more widespread distribution of
cooler, but nevertheless dense, molecular clumps.
}
{}


\maketitle

\section{Introduction}

DR21 is a deeply embedded \HII{} region created by
the radiation from at least six OB stars \citep{Roelfsema}.
It sits within a ridge of dense molecular material that obscures
the \HII{}-region at optical wavelengths. The embedded cluster drives a
violent bipolar outflow in north-east to south-west direction.
It is prominent in the  2~$\mu$m emission of vibrationally excited H$_2$,
tracing hot, shocked gas and in the Spitzer 4.5~$\mu$m channel
\citep{Garden,Davis}.
As the cluster is located close to the eastern edge of the 
molecular ridge, the eastern, blue-shifted outflow expands in 
a blister-like fountain, while the western, red-shifted outflow
is highly collimated \citep{Lane}.

Spitzer 8~$\mu$m images reveal spots of bright PAH emission
with a size below 10$''$ \citep{Marston}. They represent the
surfaces of high-density, UV irradiated clumps, forming 
photon-dominated (or photo-dissociation) regions (PDRs), 
transition zones from ionized and 
atomic gas to dense molecular gas where physics and chemistry
are dominated by UV radiation from young stars \citep{HT1999,MolSpa}.
These hot and dense regions give rise to the
emission of PDR tracers, such as HCO$^+$, high-$J$ CO,
atomic and ionized carbon, and atomic oxygen. 

An overview on the numerous existing observations in DR21 is given by 
\citet{Schneider2006} and \citet{Jakob}. {\changed
In spite of the wealth of data, the heating of the molecular
gas is still debated. The bright H$_2$ emission indicates
shock heating in the outflow and the wings of the
molecular lines prove its dynamical impact, but
excitation models} of the observed
emission of CO, \CI{}, \CII{}, and \OI{} by \citet{Lane} and
\citet{Jakob} have shown that the emission of those tracers
cannot be explained by shocks, but is consistent with a pure
UV heating, i.e., PDR physics.

To {\changed quantify the heating of the gas,} 
we analyze spectra taken with
the {\it HIFI} instrument \citep{HIFI} on board the {\it Herschel Space 
Observatory} \citep{Herschel} during the performance verification 
campaign. 
Observations of high-$J$ HCO$^+$ transitions trace hot material, 
ionized by UV radiation or X-rays \citep{Sternberg1995}.
Hot water lines are produced in shocked gas 
\citep{Snell2005}, and observations of
CO isotopes around $J=10$ close the gap in the excitation ladder
between ground-based observations and existing ISO data, allowing
to obtain a full picture of the temperature distribution.

In Sect. 2 we present the observational data. Sect. 3 compares the
measured line profiles to distinguish different components based
on their velocity distribution. In Sect. 4 we provide a model for
the emission, supporting the PDR character of the source, and discuss the results
in Sect. 5.

\section{Observations}

\subsection{HIFI observations}

All spectra presented here were obtained in performance
verification observations for the HIFI instrument. As their main goal
was to demonstrate the functionality and performance of the different
observing modes, the spectra were taken with a large variety of
observing modes and strategies.
Consequently, every spectrum was taken in a slightly different manner.
All observational parameters are summarized in Table~\ref{tab_observation}.

\begin{table}
\caption{Summary of the used HIFI observational data}
\label{tab_observation}
\begin{center}
\begin{tabular}{lrrlrr}
\hline
transition & $\nu\sub{line}$\hspace*{0.5cm} & \hspace*{-0.5cm}HPBW & observing mode$^1$ & \hspace*{-0.5cm}$t\sub{int,source}$ &
rms \\
& $[$GHz$ $]\hspace*{0.3cm} & $''$ & & $[$s$]$ & $[$K$]$ \\
\hline
HCO$^+$ 6-5 & 535.062 & 40 &  OTF map & 16 & 0.04 \\
HCO$^+$ 12-11 & 1069.694 & 21 & FSW spectral scan & 150 & 0.1 \\
$^{13}$CO 10-9 & 1101.350 & 21 & LC spectral scan & 270 & 0.08 \\
C$^{18}$O 10-9 & 1097.163 & 21 & LC spectral scan & 270 & 0.08 \\
\CII{} & 1900.537 & 12 & DBS raster map & 14 & 1 \\
& & 20 & convolved+binned & 112 & 0.2 \\
& & 40 & convolved+binned & 420 & 0.1 \\
\hline
\end{tabular}
\end{center}
$^1$ DBS = dual-beam-switch, OTF = On-The-Fly, FSW = frequency-switch, 
LC = load-chop, OFF position = 20h37m10s, 42$^\circ$37$'$00$''$
\end{table}

Most observations where single-point observations towards the
central position of the DR21 \HII{} region at RA=20h39m01.1s,
DEC=42$^\circ$19$'$43.0$''$ (J2000). Fully sampled maps were only
obtained in the [C{\sc II}] line. Data were taken with
the wideband spectrometer (WBS) at 
a resolution of 1.1~MHz, corresponding to 0.2~km/s (at 1900~GHz) --
0.7~km/s (at 500~GHz). The [C{\sc II}] data were rebinned to a 
velocity resolution of 0.45~km/s to improve the signal to noise.

\subsection{Complementary data}

ISO Long Wavelength Spectrometer $43-197$ $\mu$m grating scans
were obtained for the DR21 central position from the ISO Data Archive
(TDT 15200786).
Integrated line intensities were extracted for \OI{} at 63 and 
145\,$\mu$m and the CO 14--13 to 17--16 transitions.
%
Mid-$J$ CO lines of the DR21 region were mapped with the KOSMA 3~m 
submm telescope \citep{Jakob}. We use the lines of CO and $^{13}$CO
from $J=3$-2 to 7-6, which have been observed at native angular 
resolutions from 80$''$ to 40$''$.
The HCO$^+$ 1-0, H$^{13}$CO$^+$ 1-0, and HCO$^+$ 3-2 observations 
were taken with the IRAM 30~m telescope \citep{Schneider}. Native angular
resolutions at the 1-0 and 3-2 transition frequencies are 28$''$ and
9$''$, respectively.

\subsection{Beam size effects}
\label{sect_beam}

For a direct comparison of the different data sets we smoothed
the available data to the coarsest common angular resolution of 80$''$,
matching that of the CO 3-2 KOSMA beam and the lowest frequency
ISO observations.
This is impossible for the single-point HIFI observations. 
Moreover, the {\it Herschel} beam varies between 40$''$ HPBW for the 
HCO$^+$ line at 535~GHz and 20$''$ for $^{13}$CO at 1100~GHz. 
Those data were corrected for the different beam filling by estimating
the source size from the HCO$^+$ 3-2 line as a PDR tracer \citep{Sternberg1995}
with quite compact emission (20$''$ to 30$''$ in diameter). 
Successively convolving from the native angular 
resolution of 9$''$ to the angular resolution of the {\it Herschel} data,
and further to the final, smoothed spectra at 80$''$, 
we obtained scaling factors for the HIFI spectra, being 0.5 when going 
from 40$''$ to 80$''$ and 0.33 when going from 20$''$ to 80$''$. 
The maps obtained in [C{\sc II}] allow a direct smoothing to a resolution
of 40$''$ (see Table~\ref{tab_observation}). Beyond that size, the
same scaling factor as above was applied.
%
%

\section{Line profiles}
\label{sect_profiles}

The molecular ridge including the DR21 \HII{} region has an intrinsic
LSR velocity of -3~km~s$^{-1}$.  The quiescent material is visible in
narrow absorption lines of NH$_3$ \citep{Matsakis} and H$_2$CO
\citep{Bieging}. A second velocity component at 8-10~\kms{} is known
to be associated with the W75N complex.
It appears in emission in CO and \CI{} \citep{Jakob}, as a narrow absorption
feature in HCO$^+$ 1-0 \citep{Nyman}, and has a very broad
velocity distribution in the \HI{} 21~cm absorption \citep{Thompson,
Roberts}. The wings of the low-$J$ CO lines trace outflow velocities
down to about $-$20~\kms{} for the eastern, blister outflow and up to
$\approx 10$~\kms{} for the western outflow.

\begin{figure}
\centering
\includegraphics[angle=90.0,width=\columnwidth]{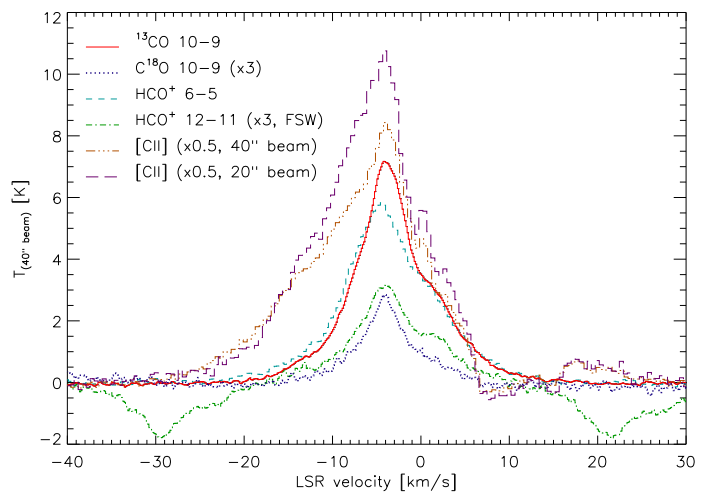}
\caption{HIFI spectra of the lines used in the fit of the PDR properties.
The \CII{} data are convolved either to the resolution of the 
HCO$^+$ 6-5 line or to that of the CO 10-9 lines. The negative features
in the HCO$^+$ 12-11 line are artifacts from the frequency-switch
observing mode.}
\label{allhifi}
\end{figure}

Figure \ref{allhifi} shows the profiles of the HIFI spectra of CO
isotopes, HCO$^+$ and \CII{}.  All lines peak at about -4~km s$^{-1}$.
The CO and HCO$^+$ lines have a similar shape, but the \CII{} line
shows an additional broad blue wing extending down to -30~km s$^{-1}$.
This indicates that the warm molecular material is slightly blue-shifted 
relative to the cold gas and that the \CII{} emission is not only 
originating from that warm gas, but also 
from the ionized wind in the blister outflow.

\begin{figure}
\includegraphics[angle=90.0,width=8.86cm]{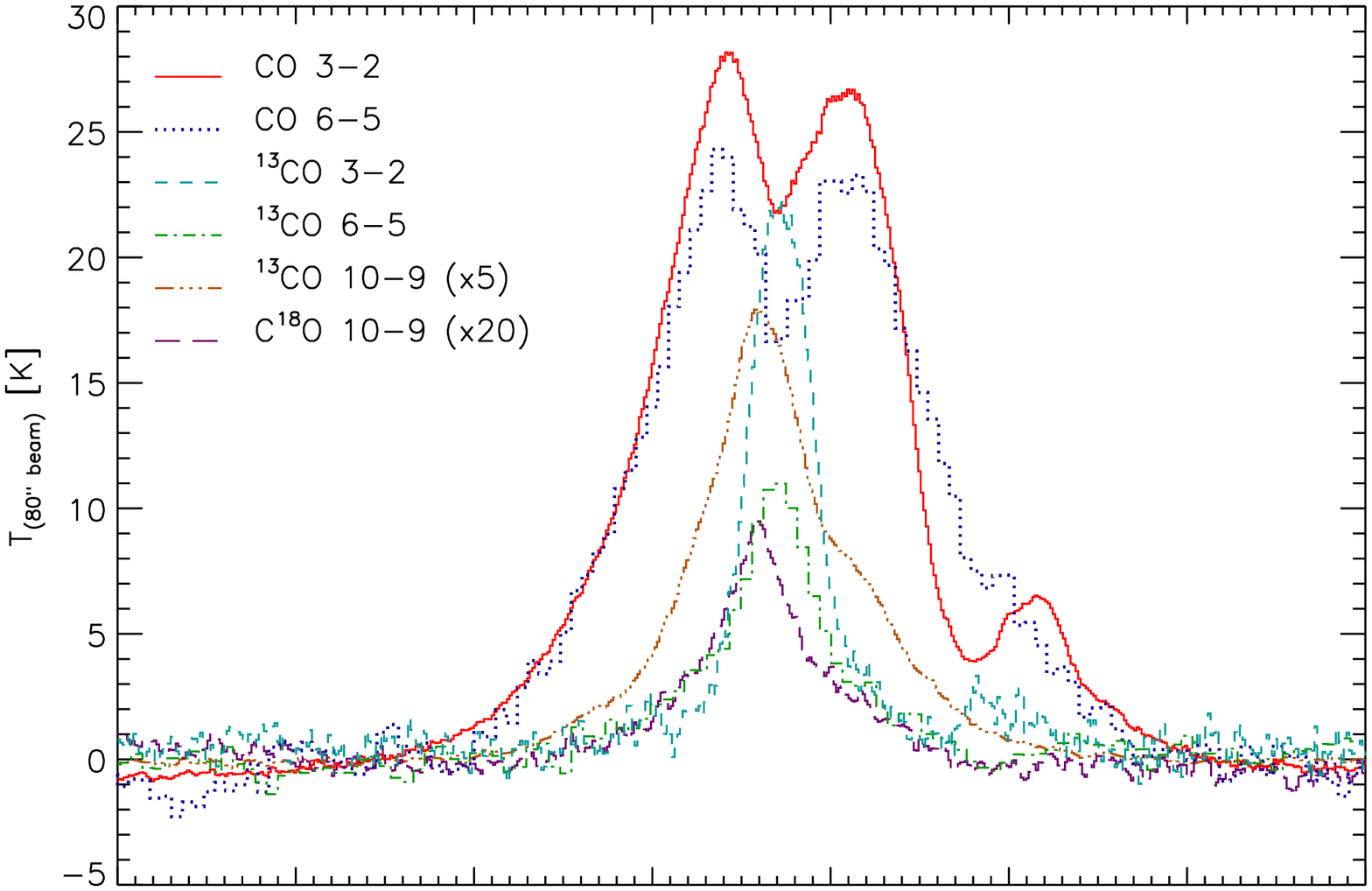}\\
\includegraphics[angle=90.0,width=\columnwidth]{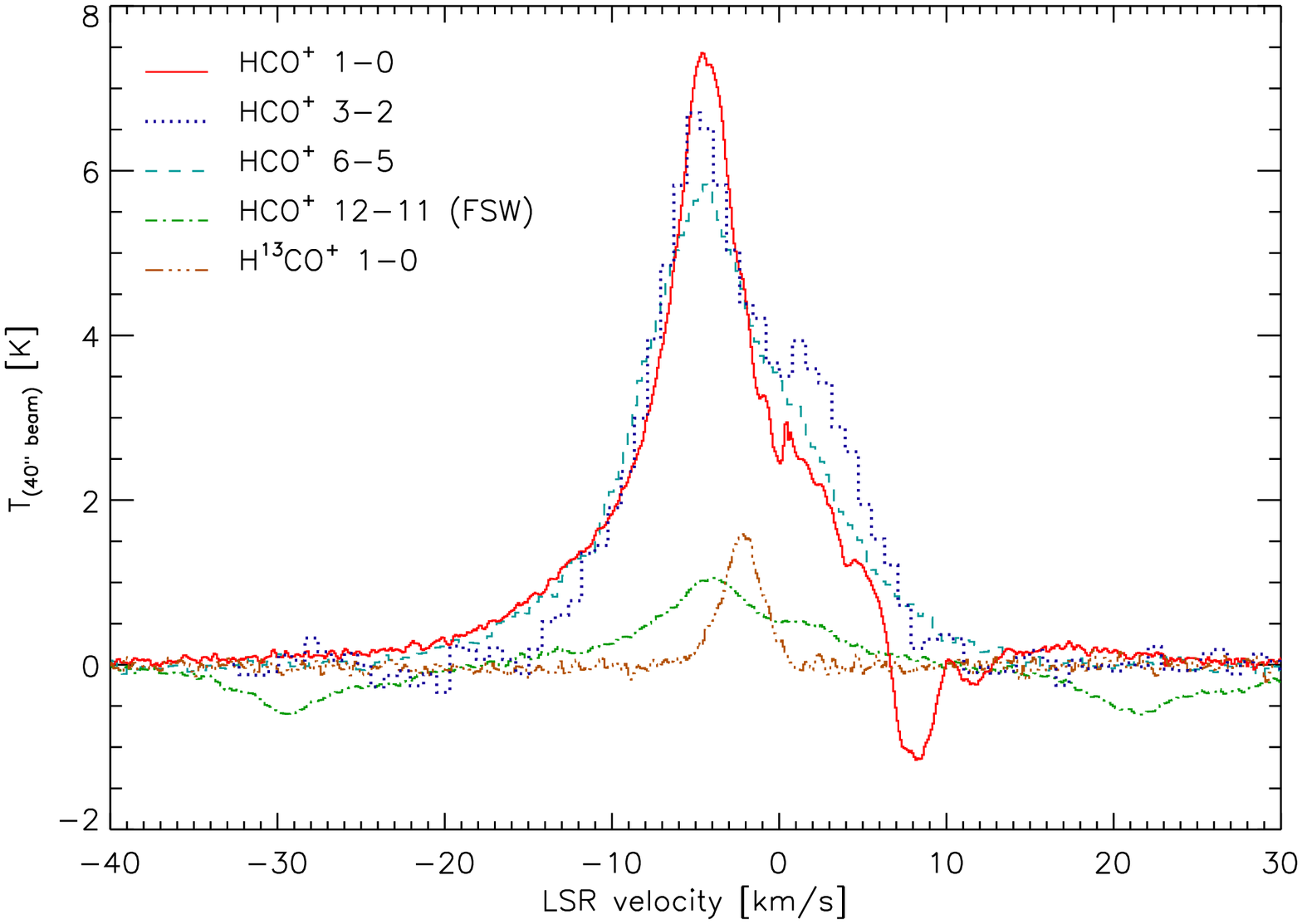}
\caption{Selected profiles of CO isotope lines (upper plot) and
HCO$^+$ isotope lines towards the DR21 central position. All
lines are convolved or scaled to the coarsest common resolution
(see Sect.\ref{sect_beam}).}
\label{fig_cohcoplus}
\end{figure}

Figure~\ref{fig_cohcoplus} compares the shapes of the CO and HCO$^+$
isotope lines with complementary ground-based measurements towards the same
positions. We show only a few selected transitions as, e.g., the
data for the CO 7-6 or 4-3 lines provide no additional 
information. All CO isotopic lines up to 7-6 are
roughly symmetric, centered at the ridge velocity of $-3$~\kms{}.
The lines of the main isotope are heavily self-absorbed with the
absorption dip marking line center. The 10-9 lines, tracing 
hotter material, are slightly asymmetric and shifted to -4~\kms{}.
All HCO$^+$ lines, except H$^{13}$CO$^+$ 1-0, have a
profile very similar to the 10-9 lines of the CO isotopes indicating
that HCO$^+$ is mainly abundant in the heated layer and less in
the overall molecular material. The H$^{13}$CO$^+$ 1-0 line
has a deviating profile centered at -2~\kms{} tracing global
infall \citep{Kirby}. CO 3-2 and HCO$^+$ 1-0 show an additional 
feature at the W75N complex velocity.

\begin{figure}
\centering
\includegraphics[angle=90.0,width=\columnwidth]{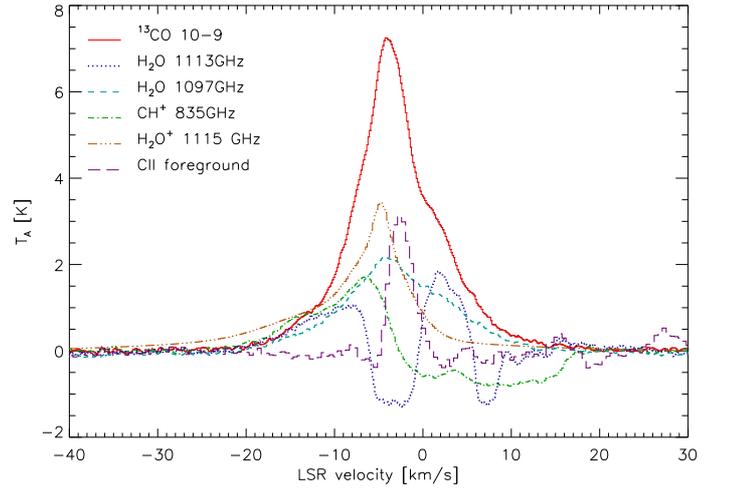}
\caption{Auxiliary lines measured by HIFI towards the same position.
They can be used to estimate the contribution of the foreground
material. The $^{13}$CO 10-9 profile is also displayed for comparison.}
\label{fig_aux}
\end{figure}

To better understand the exact velocity distribution in the source,
we plot some additional HIFI lines towards the same position 
\citep[see][]{Falgarone,WADI2,vdT} in Fig. \ref{fig_aux}. The 
1097~GHz line of hot water matches the rest of the PDR tracers.
The foreground at -3~\kms{} is
visible in absorption in the ground state transitions of water,
but in emission in a nearby OFF position for \CII{}. At
velocities of about +2~km~s$^{-1}$, we find the warm gas facing the
redshifted western outflow, apparent as a secondary peak in the ground 
state water and CH$^+$ lines and as a shoulder in $^{13}$CO and the
hot water line.
Finally, we can clearly identify the cold W75N component at $\approx -9$~\kms{}.

For the warm gas we can distinguish two velocity components -- a blue 
shifted one related to the blister outflow at $-4$~\kms{} and a second one,
at 2~\kms{}, related to the western outflow. This is consistent with 
the clumpy PDR geometry proposed by \citet{Lane}. In terms of line modelling,
it is, however, impossible to separate the two outflow directions
as their emission is overlapping to a large degree. We simply add their
intensities. The \CII{} emission requires a special treatment because of
the additional emission from the ionized outflow. To take
this into account, we have obtained an integrated \CII{} intensity
by scaling the $^{13}$CO  10-9 line profile to match the
peak and the red wing of the \CII{}  profile and ignoring the remaining
blue-wing emission.

\section{Modelling}

We use the KOSMA-$\tau$ PDR code \citep{Roellig2006} to model the 
emission of PDR ensembles, representing a distribution of spherical clumps with
$dN/dM \propto M^{-1.8}$ \citep{Cubick}. For DR21 two ensembles with
different properties had to be superimposed, a hot component, close to the 
inner \HII{} region with strong FUV illumination,
 but only a small fraction of the total mass, 
and a cooler component that fills a larger solid angle and provides
the bulk of the material. Each clumpy PDR ensemble has five free parameters:
the average ensemble density, $n_\mathrm{ens}$, the ensemble mass,
$M_\mathrm{ens}$, the UV field strength, $\chi$ given in units of
the Draine field, and the minimum and maximum mass of the clump 
ensemble, $[M_\mathrm{min},M_\mathrm{max}]$. In contrast to most
other PDR models, we fit absolute line intensities, using
the available ground-based observations, complementary ISO data, and the HIFI 
lines of the CO isotopes, HCO$^+$, atomic and ionized carbon, and atomic
oxygen.
The chemical network that has been applied in these calculations
includes $^{13}$C but not $^{18}$O. The C$^{18}$O lines 
were scaled from the $^{13}$CO intensities with a conversion of 1:8. 
Simulated annealing was used to find the optimum parameter combination.

The significance of the model is limited by the fact that the 
clump superposition ignores mutual line shading between different
clumps, i.e., optical
depth effects are only considered within individual clumps. This is
usually justified by the virialised velocity dispersion between different
clumps, but for optically very thick and broad lines some correction is
needed. To estimate the effect we have computed the optical depth for
the bulk of the individual clumps. This is of the order of unity or below
for the majority of the observed transitions, reaches values up to ten
for $^{13}$CO and HCO$^+$ transitions up to $J=4$, but exceeds ten for
the CO main isotope lines up to $J=6$ and the \OI{} 63~$\mu$m
line. For the CO lines showing clear self-absorption dips, we performed
a Gaussian fit to the line wings and used the integrated intensity of
that Gaussian to compute the total emission including the blocked
radiation from the inner clumps close to the \HII-region. As
we have no spectral information for the \OI{} line, we have no
estimate for the blocked radiation in this case, so that we excluded
that data point from the fit.

\begin{figure}
\centering
\includegraphics[angle=0.0,width=\columnwidth]{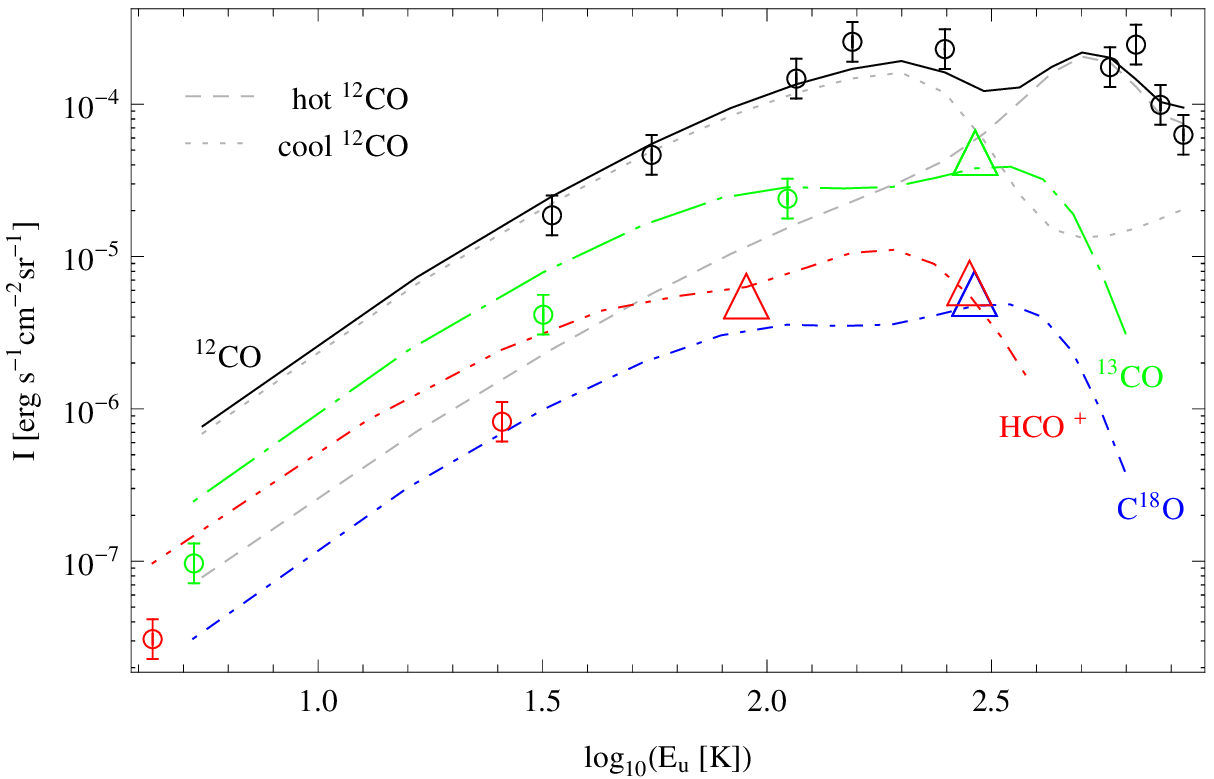}\\
\includegraphics[angle=0.0,width=\columnwidth]{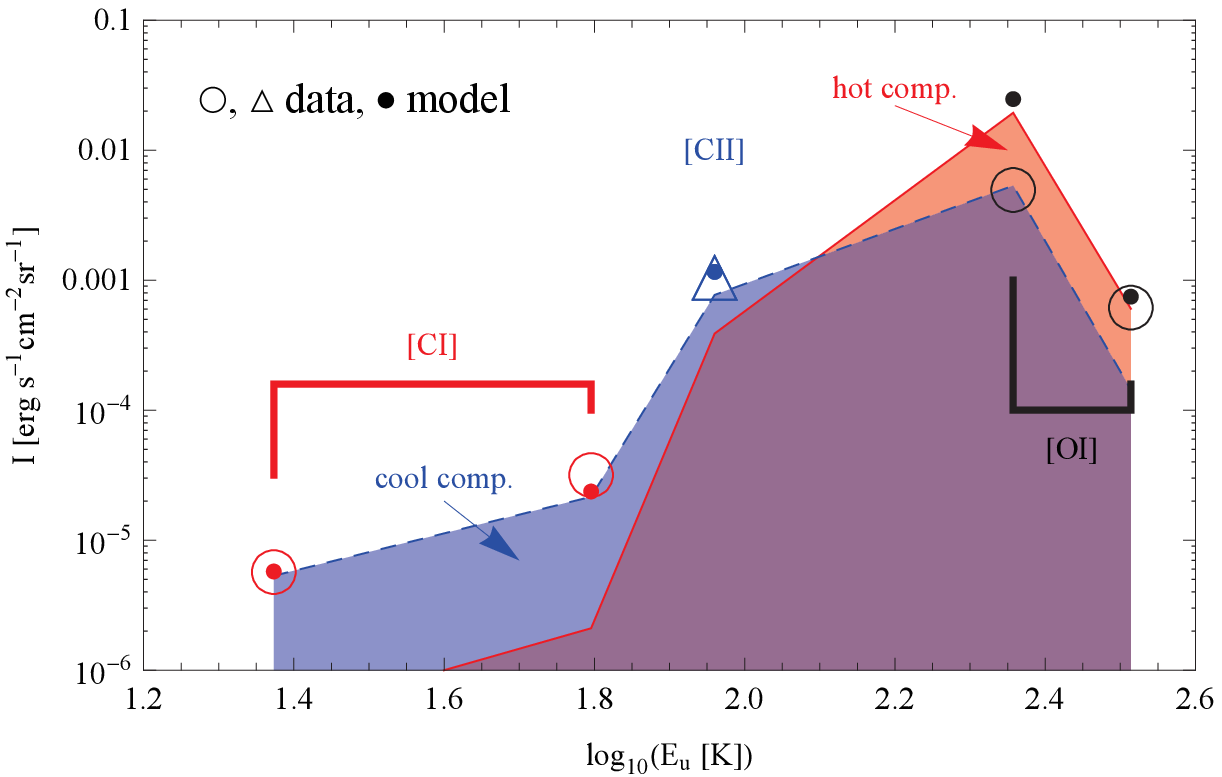}
\caption{Two-ensemble PDR model fit to the observed CO, HCO$^+$, and
fine structure line intensities, shown as function of the upper
level energy. HIFI measurements are depicted as open triangles, 
complementary data points as open circles. 
The dotted line and the shaded areas indicate the contributions
from the two ensembles to the CO and the fine structure lines, respectively.}
\label{fig_modelfit}
\end{figure}

The best fit result is shown in Figure~\ref{fig_modelfit}. The
corresponding model parameters are:
\vspace*{-0.3cm}
\begin{description}
\item[Ensemble 1:]$\chi=1.0\,10^{5}$, $n\sub{ens}=1.3\,10^{6}$cm$^{-3}$, $M\sub{ens}=150~M_\odot$, $[M\sub{min},M\sub{max}]=[10^{-2},8\,10^1] M_\odot$
\item[Ensemble 2:]$\chi=3.2\,10^{2}$, $n\sub{ens}=1.1\,10^{6}$cm$^{-3}$, $M\sub{ens}=830~M_\odot$, $[M\sub{min},M\sub{max}]=[10^{-3},10^1] M_\odot$
\end{description}
\vspace*{-0.3cm}
The two-ensemble model provides a reasonable fit to all 22 data points.
The model slightly overestimates the emission for the optically thick
rotational lines of {\changed $^{13}$CO and HCO$^+$ which showed no
clear self-absorption dip, so that we have insufficient optical depth
corrections. The reduced $\chi^2$ value of the fit amounts to 11.1, where
a contribution of 9.5 results from the three optically thick low-$J$
data points of $^{13}$CO and HCO$^+$. For the remaining 19 data points
we would obtain an excellent reduced $\chi^2$ value of 1.6. The
topology of the $\chi^2$ function shows several side minima, but they are 
worse by at least a factor two.} As expected,
the model predicts a too high \OI{} intensity, as it ignores that
the outer clumps of the cooler ensemble block the contribution from
the hot inner component. 

\section{Discussion}

The double-peak structure of the predicted CO intensities 
across the excitation ladder reflects the two different UV fields
leading to different excitation conditions. 
No single-parameter ensemble can fit all data, but distinct low
temperature and high temperature components are needed.
The parameters of the model
are in agreement with independent estimates. A UV flux of $10^5$ Draine fields
corresponds to a geometrical distance of 0.06~pc, i.e. 7$''$, from the central
cluster, matching the size of the PAH emission \citep{Marston}. Our clump
densities match those determined by \citet{Jakob} for the extended cool
gas, but are slightly higher than their hot-gas density
($4\,10^5$~cm$^{-3}$). In contrast, \citet{Jones, vdT} find
still somewhat higher densities for the hot gas, up to $10^7$~cm$^{-3}$.
The hot ensemble mass is close to the 170~$M_\odot$ derived from 
early CO 7-6 observations by \citet{Jaffe1989}.
The total mass of the PDR ensembles falls between the mass limits 
derived by \citet{Jakob}
from dust observations and from line radiative transfer fits. 

While the existing ground-based observations provide a very good constraint
on the properties of the extended cool gas, and the ISO lines show the
total amount of hot gas, it is only the set of new HIFI data that
puts the hot and cold distributions well apart from each other in terms of
the temperature structure. While \citet{Jakob} obtained 
cooling curves with single peaks, the new data for the 10-9 lines of
the CO isotopes and the HCO$^+$ transitions force the fit to a bimodal 
distribution of excitation conditions. When we exclude the {\it Herschel} data
from the model fit, we obtain a parameter set that shows a UV field that
is lower by a factor ten for the hot ensemble, i.e., that would imply
molecular clumps farther away from the central cluster.
For the cold ensemble, the fitted UV field is also somewhat lower, while all
other parameters remain similar to those from the full fit. Only with
the {\it Herschel} data, we therefore obtain a parameter set that is consistent
with the source geometry.

As the two-ensemble PDR model is able to fit all of the observed lines, we 
find no evidence for a shock heating of the dense gas. This is in agreement
with the analysis of \citet{Lane}, explicitely excluding a shock origin of
the fine-structure lines,
but seems to be in contradiction with the analysis of the line profiles in
Sect. \ref{sect_profiles} that shows excited outflow material. We conclude
that the material visible in the blue line wing, characterizing the blister
outflow, is contained in dense clumps that are accelerated by the outflow,
but that are chemically and energetically fully dominated by the UV field
and not by the associated shock.

\begin{acknowledgements}
HIFI has been designed and built by a consortium of institutes and university departments from across
Europe, Canada and the United States under the leadership of SRON Netherlands Institute for Space
Research, Groningen, The Netherlands and with major contributions from Germany, France and the US.
Consortium members are: Canada: CSA, U.Waterloo; France: CESR, LAB, LERMA, IRAM; Germany:
KOSMA, MPIfR, MPS; Ireland: NUI Maynooth; Italy: ASI, IFSI-INAF, Osservatorio Astrofisico di Arcetri-
INAF; Netherlands: SRON, TUD; Poland: CAMK, CBK; Spain: Observatorio Astronómico Nacional (IGN),
Centro de Astrobiolog\'ia (CSIC-INTA). Sweden: Chalmers University of Technology - MC2, RSS \& GARD;
Onsala Space Observatory; Swedish National Space Board, Stockholm University - Stockholm Observatory;
Switzerland: ETH Zurich, FHNW; USA: Caltech, JPL, NHSC.

      This work was supported by the German
      \emph{Deut\-sche For\-schungs\-ge\-mein\-schaft, DFG\/} project
      number Os~177/1--1. 
A portion of this research was performed at the Jet Propulsion Laboratory, California Institute of Technology, under contract with the National Aeronautics and Space Administration.
\end{acknowledgements}

\end{document}